\documentclass[12pt]{iopart}

\usepackage{amsfonts, amssymb, mathrsfs, times, float, color, bm}
\usepackage{graphicx}
\usepackage{graphicx,epstopdf}
\usepackage{dcolumn}
\usepackage{bm}
\newcommand{\nn}{\nonumber}

\begin{document}

\title[Post-Keplerian motion]{Post-Keplerian motion in Reissner-Nordstr\"{o}m spacetime}

\author{Bo Yang$^{1}$ \& Wenbin Lin$^{2,1,\ast}$}

\address{$^{1}$School of Physical Science and Technology, Southwest Jiaotong University, Chengdu 610031, China \\$^{2}$School of Mathematics and Physics, University of South China, Hengyang, 421001, China}

\ead{lwb@usc.edu.cn}
\vspace{10pt}
%\begin{indented}
%\item[]January 2016
%\end{indented}

\begin{abstract}
We present the analytical post-Newtonian solutions for the test particle's motion in the Reissner-Nordstr\"{o}m spacetime. The solutions are formulated in the Wagoner-Will representation, the Epstein-Haugan representation, the Brumberg representation, and the Damour-Deruelle representation, respectively. The relations between the (post-)Keplerian parameters in different representations, as well as their relations to the orbital energy and angular momentum are also provided.
\end{abstract}

%\pacs{}

\vspace{1pc}
\noindent{\it Keywords}: post-Minkowskian motion, post-Newtonian approximations, Reisser-Nordstr\"{o}m spacetime

\section{Introduction} \label{sec:intro}

The analytical solutions for the test particle's motions are of importance in both the theoretical significance and in the real applications, and have been occupied a very important status in the gravitation theories. For the Newton theory, the particle's motions in the spherically symmetric field can be described by the Kepler solutions perfectly. The analytical solution for the motion of test particle in Schwarzschild spacetime is first achieved in terms of elliptic functions by Hagihara~\cite{Hagihara1931}. Since then, the analytical solutions for the motion of the (neutral or charged) test particles have also been explored for the spacetimes of the other classical black holes, e.g, the Reissner-Nordstr\"{o}m spacetime~\cite{Gackstatter1983,GrunauKagramanova2011}, the Kerr spacetime~\cite{Mino2003,FujitaHikida2009}, and the Kerr-Newman spacetime~\cite{HackmannXu2013}. All these solutions are formulated in terms of different elliptic functions.

The solutions formulated with the elliptic functions are exact and concise, but the elliptic functions are not convenient in deriving the quantities related to the motion such as orbital periods and gravitational waveforms, and can not explicitly exhibit the effects of the gravitational source's parameters on the motion of the test particle. For the gravitational fields being not too strong, we can employ the post-Newtonian (PN) approximations to study the test particle's motions in these kinds of fields, and the motions of the test particle can be formulated without involving the special functions.
The analytical PN solutions for the test particle's motion and gravitational radiation in the Kerr and Lense-Thrring spacetimes have been studied~\cite{Shibata1994,Ryan1996,GPV1998a}. %{,Ryan1995,Ryan1996,GPV1998a}.
At the other hand, since the binary systems are very important in testing the gravitation theories, a number of analytical solutions for the binary systems have also been achieved, not only for the first PN order~\cite{Brumberg1972,WagonerWill1976,Epstein1977,DamourDeruelle1985,Haugan1985,Soffel1987,Soffel1989,KlionerKopeikin1994,KopeikinEfroimskyKaplan2012,PoissonWill2014}, but also for the higher PN orders for the mass~\cite{DamourSchafer1988,SchaferWex1993,MoraWill2004,MGS2004,Boetzel2017}, as well as the spin's effects~\cite{Wex1995,KonigsdorfferGopakumar2005,KMG2005,THS2010,GopakumarSchafer2011,BMFB2013,GergelyKeresztes2015,Mikoczi2017}, and the quadrupole's effects~\cite{Poisson1998,BiniGeralico2014}. The solutions for the test particle's motion can be obtained from those for the binary systems directly via taking the limit of extreme mass ratio.

In this work, we study the PN solutions for the test particle's motion in the Reissner-Nordstr\"{o}m spacetime, which characterizes the field of a classical charged black hole. We derive the post-Keplerian orbit and period in the harmonic coordinates, and the solutions are formulated in the Wagnoner representation~\cite{WagonerWill1976}, the Epstein-Haugan representation~\cite{Epstein1977,Haugan1985}, the Brumberg representation~\cite{Brumberg1972} and the Damour-Deruelle representation~\cite{DamourDeruelle1985}, respectively.

The rest of this paper is organized as follows. Section \ref{sec:2nd} gives the post-Newtonian dynamics for the test particle in the Reissner-Nordstr\"{o}m spacetime. In Section \ref{sec:3rd} we present the post-Keplerian solutions for the orbit and period. Section \ref{sec:4rd} gives the relations between the (post-)Keplerian parameters and the orbital energy and angular momentum. The summary is given in Section \ref{sec:5rd}.

\vskip 0.5cm

\section{The post-Newtonian dynamics for the test particle}\label{sec:2nd}
In the harmonic coordinates, the metric of Reissner-Nordstr\"{o}m black hole in the first PN approximation can be written as~\cite{LinJiang2014}
\begin{eqnarray}
&& g_{00}= -1 +\frac{2M}{r} -\frac{2m^2+q^2}{r^2}~,\label{eq:metric-1nd}\\
&& g_{0i}= 0~,\label{eq:metric-2nd}\\
&& g_{ij}=\Big(1\!+\!\frac{2 m}{r}  \Big)\delta_{ij} ,\label{eq:metric-3nd}
\end{eqnarray}
where $m$ and $q$ denote the mass and electric charge of the black hole. $r\!\equiv\! |\bm{x}|$ denotes the distance from the field position $\bm{x}\!\equiv\!(x^1,x^2,x^3)$ to the black hole located at the coordinate origin. The gravitational constant and the light speed in vacuum are set as $1$. The metric has signature of ($-+++$). Latin indices $i$ and $j$ run from 1 to 3. The corresponding Lagrangian reads:
\begin{eqnarray}
&& {\rm L} = \frac{1}{2}\bm{v}^{2}+\frac{m}{r}+\frac{1}{8}\bm{v}^{4}+\frac{3}{2}\frac{m}{r}\bm{v}^{2}-\frac{1}{2}\frac{m^2\!+\!q^2}{r^2}~.
\end{eqnarray}

The orbital energy $\mathcal{E}$ and the normalized orbital angular momentum $\mathcal{J}$ in the Reissner-Nordstr\"{o}m spacetime can be calculated by %as follows:
\begin{eqnarray}
&& \mathcal {E}=%\bm{v}\cdot\frac{\partial\rm{L}}{\partial\bm{v}}-{\rm L}=
\frac{1}{2}\bm{v}^{2}-\frac{m}{r}+\frac{3}{8}\bm{v}^{4}+\frac{3}{2}\frac{m}{r}\bm{v}^{2}+\frac{1}{2}\frac{m^2\!+\!q^2}{r^2}~,\label{orbit-energy}\\
&& \hskip 0.4cm \mathcal{J}%=|\bm{x}\times\frac{\partial\rm{L}}{\partial\bm{v}}|
=\frac{|\bm{x}\times\bm{v}|}{m}\Big(1+\frac{1}{2}\bm{v}^{2}+\frac{3m}{r}\Big)~,\label{orbit-angularmomentum}
\end{eqnarray}
where $\bm{v}\!\equiv\! \frac{d\bm{x}}{dt}$ denotes the velocity of the test particle.

The motion of the test particle is described by the geodesic equation. Based on Eq.\,(\ref{eq:metric-1nd})-(\ref{eq:metric-3nd}), we can obtain the equations of motion of the particle
\begin{eqnarray}
\frac{d^2\bm{x}}{dt^2} = -\frac{m\bm{x}}{r^3}\Big[1\!-\!\frac{m}{r}\Big(4+\frac{q^2}{m^2}\Big)+v^2\Big]\!+\!\frac{4m(\bm{v}\cdot\bm{x})}{r^3}\bm{v}~.\label{acceleration}
\end{eqnarray}

In the next section we will derive the post-Keplerian solutions to Eq.\,(\ref{acceleration}) via several classical representations.

\vskip 0.5cm

\section{The post-Keplerian motion in Reissner-Nordstr\"{o}m spacetime}\label{sec:3rd}

Since the Kepler solutions are the basis to derive the post-Keplerian solution for the particle's motion, we include it here for the completeness.

\subsection{The Kepler solutions}
In the frame of Newtonian theory, the charge of gravitational source does not have effects on the particle's motion, and we assume that the particle moves along the Kepler orbit with an eccentricity $e$ and a semi-latus rectum $p$. Without the loss of generality, we can take the plane in which the particle moves as the equatorial plane, then the particle's position vector $\bm{x}$ can be written in terms of $r$ and $\phi$
\begin{eqnarray}
\bm{x} = r(\cos\phi\,\bm{e}_{x}+\sin\phi\,\bm{e}_{y})~,
\end{eqnarray}
and the Kepler solutions gives:
\begin{eqnarray}
%&& \bm{v} = \left(\frac{m}{p}\right)^{\frac{1}{2}}\!\left[-\!\sin\phi\,\bm{e}_{x}\!+\!(e+\cos\phi)\bm{e}_{y}\right]\\
&& \hskip 2cm r^2\frac{d\phi}{dt} = (mp)^{\frac{1}{2}}~, \\
&& \bm{v} = \Big(\frac{m}{p}\Big)^{\frac{1}{2}}\!\left[-\!\sin\phi\,\bm{e}_{x}\!+\!(e+\cos\phi)\bm{e}_{y}\right]~,\\
&& \hskip 1.9cm  r = \frac{p}{1+e\cos\phi}~,\\
&& \hskip 1.3cm  t\,\Big(\frac{2\pi}{{\rm T}_{E}}\Big) = E-e\sin E~.
\end{eqnarray}
where $E$ denotes the eccentric anomaly, and ${\rm T}_{E}$ denotes the period for the eccentric anomaly
\begin{eqnarray}
&& {\rm T}_{E} = 2\pi\Big(\frac{a^3}{m}\Big)^{\!\frac{1}{2}}~,\label{PeriodK}
\end{eqnarray}
with $a \!\equiv \!p/(1-e^2)$ being the Kepler orbital semimajor axis. The relation between $\phi$ and $E$ is
\begin{eqnarray}
&&  \sin\phi=\frac{(1-e^2)^{\frac{1}{2}}\sin E}{1-e\cos E}~,~~~~~~\cos\phi=\frac{\cos E-e}{1-e\cos E}~.~\label{phi-E-relation}
\end{eqnarray}
The Kepler orbital period can be described by ${\rm T}_{E}$.

\subsection{The Wagoner-Will representation}\label{Wagoner-Will representation}
Following the method given by Wagoner and Will~\cite{WagonerWill1976}, we can obtain the PN solutions for the particle's motion in Reissner-Nordstr\"{o}m spacetime as follows
\begin{eqnarray}
&&   r^2\frac{d\phi}{dt} = (mp)^{\frac{1}{2}}\Big(1-\frac{4me}{p}\cos\phi\Big)~,\label{dphidt}
\end{eqnarray}
\begin{eqnarray}
&& \hskip -2cm \bm{v} = \Big(\frac{m}{p}\Big)^{\!\frac{1}{2}}\!\Big[-\!\sin\phi\,\bm{e}_{x}\!+\!(e+\cos\phi)\bm{e}_{y}\Big]\nn\\
&&\hskip -1.75cm +\bm{e}_{x}\Big\{\Big(\frac{m}{p}\Big)^{\!\frac{3}{2}}\Big[\!-\!3e\phi\!+\!3\sin\phi\!-\!e^2\sin\phi\!+\!\frac{1}{2}e\sin2\phi
\!+\!\frac{q^2}{m^2}\Big(\frac{1}{2}e \phi\!+\!\sin\phi \!+\! \frac{1}{4}e\sin2\phi\Big)\Big]\Big\}\nn\\
&&\hskip -1.75cm  -\bm{e}_{y}\Big\{\Big(\frac{m}{p}\Big)^{\!\frac{3}{2}}
\Big[3\cos\phi\!+\!3e^2\!\cos\phi\!+\!\frac{1}{2}e\cos2\phi\!+\!\frac{q^2}{m^2}\Big(\cos\phi \!+\! \frac{1}{4}e\cos2\phi\Big)\Big]\Big\},
\end{eqnarray}
\begin{eqnarray}
&& \hskip -2cm \frac{p}{r}= 1\!+\!e\cos\phi\!+\!\frac{m}{p}\Big[\!-\!3\!+\!e^2\!+\!\frac{7}{2}e\cos\phi
\!+\!3e\phi\sin\phi\!-\!\frac{q^2}{m^2}\Big(1\!+\!\frac{1}{2}e\phi\sin\phi \!+\! \frac{1}{4}e\cos\phi\Big)\Big].\label{rphi1}
\end{eqnarray}

The orbital precession per revolution $\Delta \varphi$ can be obtained from Eq.\,(\ref{rphi1}) %and it reads
\begin{eqnarray}
 \Delta \varphi = 2\pi\Big[\frac{m}{p}\Big(3-\!\frac{q^2}{2m^2}\Big)\Big]~.\label{Deltaphi}
\end{eqnarray}

Substituting Eq.\,(\ref{rphi1}) into Eq.\,(\ref{dphidt}), and keeping all terms to the first post-Newtonian order, we can achieve the particle's orbital period as follow
\begin{eqnarray}
\hskip -2cm  {\rm T}_{\phi} \,= \int_{0}^{2\pi+\Delta \varphi} \!\frac{dt}{d\phi} d\phi  \,=\, 2\pi\Big(\frac{a^3}{m}\Big)^{\!\frac{1}{2}}\!\Big\{1\!+\!\Big[9\!+\!\frac{9}{2}e^2\!+\!3e^4\!+\!\frac{q^2}{m^2}\Big(\frac{3}{2}\!+\!\frac{3}{4}e^2\Big)\Big]\frac{m}{a(1-e^2)^2}\Big\}~.
~\label{Period1}
\end{eqnarray}

\subsection{The Epstein-Haugan representation}\label{Epstein-Haugan representation}
With the variable ``true anomaly" $\eta$, Eqs.\,(\ref{dphidt}) and  (\ref{rphi1}) can be written in a periodic form~\cite{Epstein1977,Haugan1985,Soffel1987}  %(Epstein 1977,Haugan 1985)
\begin{eqnarray}
&& \hskip 2.5cm  \eta=\Big[1-\frac{m}{p}\Big(3-\frac{q^2}{2m^2}\Big)\Big]\phi~,\\
&& \hskip 2.2cm   r^2\frac{d\phi}{dt} = (mp)^{\frac{1}{2}}\bigg[1-\frac{4me\cos\eta}{p}\bigg]~,\label{new-dphidt}\\
&& \frac{p}{r}= 1+e\cos\eta+\frac{m}{p}\Big[\!-\!3\!+\!e^2\!+\!\frac{7}{2}e\cos\eta\!-\!\frac{q^2}{m^2}\Big(1\!+\! \frac{1}{4}e\cos\eta\Big)\Big].\label{new-rphi1}
\end{eqnarray}

We can also formulate the orbit in terms of the ``eccentric anomaly" $E'$ which is related to the ``true anomaly" $\eta$ by
\begin{eqnarray}
&&  \sin\eta=\frac{(1-e^2)^{\frac{1}{2}}\sin E'}{1-e\cos E'}~,~~~~~~\cos\eta=\frac{\cos E'-e}{1-e\cos E'}~,~\label{E-relation}
\end{eqnarray}
then the orbit can be written as
\begin{eqnarray}
&& \hskip -0.2cm r= a(1\!-\!e\cos E')+\frac{m}{(1\!-\!e^2)^2}\Big\{\Big[3\!+\!\frac{23}{4}e^2\!-\!\frac{e^4}{2}\!+\!\frac{q^2}{m^2}\Big(1\!+\!\frac{e^2}{8}\Big)\Big]\!\nn\\
&& \hskip 0.5cm
-\Big[\frac{19}{2}\!+\!\frac{3}{2}e^2\!-\!\frac{q^2}{m^2}\Big(\frac{7}{4}\!-\!\frac{e^2}{4}\Big)\Big]e\cos E' +\!\Big[\frac{13}{4}\!-\!\frac{e^2}{2}\!+\!\frac{3}{8}\frac{q^2}{m^2}\Big]e^2\cos 2E'\Big\}~.\label{rE}
\end{eqnarray}

From the above relations among $\phi$, $\eta$ and $E'$, we can obtain
the dependence of the ``eccentric anomaly" $E'$ on the time $t$ as follow
%the relation between the ``eccentric anomaly" $E'$ and the time $t$ in the following form:
\begin{eqnarray}
&& t\,\Big(\frac{2\pi}{{\rm T}_{E'}}\Big) = E'-g\sin E'-h\sin 2E'~,
\end{eqnarray}
where ${\rm T}_{E'}$ denotes the $E'$-period
\begin{eqnarray}
&& {\rm T}_{E'} =  \,2\pi\Big(\frac{a^3}{m}\Big)^{\!\frac{1}{2}}\!\Big\{1\!+\!\frac{m}{a(1\!-\!e^2)^2}\Big[9\!+\!\frac{9}{2}e^2\!+\!3e^4\!+\!\frac{q^2}{m^2}\Big(\frac{3}{2}\!+\!\frac{3}{4}e^2\Big)\Big]
\Big\}~,~\label{Period2}
\end{eqnarray}
and
\begin{eqnarray}
&& g = e\Big\{1\!+\!\frac{m}{a(1\!-\!e^2)^2}\Big[9\!-\!\frac{1}{2}e^2\!-\!3e^4\!-\!\frac{q^2}{m^2}\Big(\frac{11}{2}\!-\!\frac{1}{4}e^2\Big)\Big]\Big\}~,\\
&& \hskip 1.8cm h =\frac{me^2}{a(1\!-\!e^2)^2}\Big(\!-\!\frac{13}{4}\!+\!\frac{1}{2}e^2\!-\!\frac{3}{8}\frac{q^2}{m^2}\Big)~.
\end{eqnarray}
The formulas for $g$ and $h$ will become identical to those for the two-body system in general relativity given by Klioner and Kopeikin~\cite{KlionerKopeikin1994}, when the source's charge are dropped in the former and the limit of extreme mass ratio is taken in the latter.

It can be seen that the formula for the $E'$-period given in Eq.\,(\ref{Period2}) is the same as that for the orbital period ${\rm T}_{\phi}$ given in Eq.\,(\ref{Period1}). In other words, when the angle $\phi$ goes from $0$ to $2\pi+\Delta \varphi$, the angle $E'$ goes from $0$ to $2\pi$ in the same period.

\subsection{The Brumberg representation}\label{Brumberg representation}

Following the method of Brumberg~\cite{Brumberg1972}, we can formulate the post-Newtonian solution as follows
\begin{eqnarray}
&& r=\frac{a_r(1-e_r^2)}{1+e_r\cos{f}}~,\label{rBrumberg}
\end{eqnarray}
with $f$ being the true anomaly which obeys
\begin{eqnarray}
&& f=\Big[1-\frac{m}{a_r(1-e_r^2)}\Big(3\!-\!\frac{q^2}{2m^2}\Big)\Big]\phi~.\label{dfdphi}
\end{eqnarray}
here we can see that $r_{\pm}=a_r(1\pm e_r)$ are the minimal and maximal values for the post-Keplerian orbit, thus $a_r$ and $e_r$ can be regarded as the generalized orbital semimajor axis and eccentricity of the post-Newtonian solution, and are called as the post-Keplerian parameters, in order to distinguish the Keplerian parameters ($a,\,e$). The relations between ($a_r,\,e_r$) and ($a,\,e$) are as follows
\begin{eqnarray}
&&  a_r = a\, \Big\{1+\frac{m}{a(1\!-\!e^2)^2}\Big[(3\!+\!9 e^2\!-\!e^4)+\frac{q^2}{m^2}\Big(1\!+\!\frac{1}{2}e^2\Big)\Big]\Big\}~,\label{ar}\\
&& \hskip 1.1cm
e_r = e\,\Big\{1+\frac{m}{a(1\!-\!e^2)}\Big[\Big(\frac{13}{2}\!-\!e^2\Big)+\frac{3}{4}\frac{q^2}{m^2}\Big]\Big\}~.\label{er}
\end{eqnarray}
%The orbital period can be obtained as follows.

Similarly, introducing the eccentric anomaly $E$ which is related to $f$ via
\begin{eqnarray}
&&  \sin f=\frac{(1-e_r^2)^{\frac{1}{2}}\sin E}{1-e_r\cos E}~,~~~~~~\cos f=\frac{\cos E-e_r}{1-e_r\cos E}~.~\label{trueE-relation}
\end{eqnarray}
we can obtain the relation between the eccentric anomaly $E$ and the time $t$ as follow
\begin{eqnarray}
&& t\,\Big(\frac{2\pi}{{\rm T}_{E}}\Big) = E - \Big(1-\frac{4m}{a_r}\Big)e_r \sin E ~,\label{t-E1}
\end{eqnarray}
with
\begin{eqnarray}
&& {\rm T}_{E} = 2\pi\Big(\frac{a_r^3}{m}\Big)^{\!\frac{1}{2}}\!\Big(1\!+\!\frac{9}{2}\frac{m}{a_r}\Big)~,\label{Period3}
\end{eqnarray}
representing the period of the eccentric anomaly. Later we will show that this formulation of period is identical to those in the Wagnoner-Will representation and the Epstein-Haugan representation.

\subsection{The Damour-Deruelle representation}\label{Damour-Deruelle representation}
Following the method of Damour and Deruelle~\cite{DamourDeruelle1985}, we can obtain the PN solution in the Reissner-Nordstr\"{o}m spacetime in terms of the post-Keplerian parameters ($a_r,\,e_r$) and the eccentric anomaly $E$ as
\begin{eqnarray}
&& r=a_r(1-e_r \cos E)~.\label{rDamour}
\end{eqnarray}
Notice that the definitions of $a_r$ and $e_r$ as well as the relations between the eccentric anomaly $E$ and the time $t$ are the same as those in the Brumberg representation, so we do not bother to give them here.

\vskip 0.5cm

\section{The relations between the (post-)Keplerian parameters and the orbital energy and angular momentum}\label{sec:4rd}
In the Wagnoner-Will representation and the Epstein-Haugan representation, %both of which are in terms of the Kepler parameters $(a,\,e)$,
we can calculate the orbital energy and angular momentum from Eqs.\,(\ref{orbit-energy}) and (\ref{orbit-angularmomentum}) as follows: \begin{eqnarray}
&& \mathcal{E}=-\frac{m}{2p}\Big\{(1-e^2)-\frac{m}{4p}\Big[19+22e^2+3e^4+\frac{q^2}{m^2}(4+2e^2)\Big]\Big\}~,\label{Will-total-energy}\\
&& \hskip 2.8cm \mathcal{J}=\Big(\frac{p}{m}\Big)^{\!\frac{1}{2}}\Big[1+\frac{m}{2p}\left(7+e^2\right)\Big]~.\label{Will-total-angular momentum}
\end{eqnarray}

We can also write the Keplerian parameters ($p,\,e)$ in terms of the orbital energy and angular momentum
%the energy and angular momentum in terms of the orbital parameters $e$ and $p$ as follows:
\begin{eqnarray}
&& \hskip 2.5cm p={m\mathcal{J}^2}\Big[1-\frac{8\!+\!2 \mathcal{E}\mathcal{J}^2}{\mathcal{J}^2}\Big]~,\label{p1}\\
&& e^2=1+2\mathcal{E}\mathcal{J}^2-\frac{11\!+\!30\mathcal{E} \mathcal{J}^2 \!+\!7 \mathcal{E}^2 \mathcal{J}^4}{ \mathcal{J}^2}\!-\!\frac{3\!+\!2\mathcal{E} \mathcal{J}^2}{2\mathcal{J}^2}\frac{q^2}{m^2}~.\label{e1}
\end{eqnarray}

Notice that $\mathcal{E}$ and $\mathcal{J}$ contain the contributions from both the Newtonian parts and the PN parts, therefore, when we use them to express the Keplerian parameters $p$ and $e$, we need to subtract the PN contributions from them. This is the reason that the formulas for $p$ and $e$ look like having the PN contributions from the source's mass and charge, but indeed not.

In the Brumberg representation and the Damour-Deruelle representation, we can obtain the formulas for the orbital energy and angular momentum as follows:
\begin{eqnarray}
&& \hskip 2.5cm \mathcal{E}=-\frac{m}{2a_r}\Big(1-\frac{7}{4}\frac{m}{a_r}\Big)~,\label{New-total-energy}\\
&&  \mathcal{J}=\Big[\frac{a_r(1\!-\!e_r^2)}{m}\Big]^{\frac{1}{2}}\Big\{1+\frac{m}{a_r(1\!-\!e_r^2)}
\Big[(2\!+\!e_r^2)\!-\!\frac{q^2}{2m^2}\Big]\Big\}~.\label{New-total-angular momentum}
\end{eqnarray}

Vice verse, we can write the post-Keplerian parameters ($a_r,\,e_r)$ in terms of the orbital energy and angular momentum
\begin{eqnarray}
&& \hskip 1cm a_r=-\frac{m }{2 \mathcal{E}}\Big(1+\frac{7}{2} \mathcal{E}\Big)~,\label{a2}\\
&& e_r^2=1\!+\!2 \mathcal{E} \mathcal{J}^2-\mathcal{E}(12\!+\!15 \mathcal{E} \mathcal{J}^2)+2\mathcal{E}\frac{q^2}{m^2}~.\label{e2}
\end{eqnarray}

With these relations, the orbit can be formulated in terms of the orbital energy and angular momentum, and it can be demonstrated that the solutions in all representations are identical to the first PN order. The orbit can be written as
%\begin{eqnarray}
%&&  \frac{r}{-\frac{m }{2 \mathcal{E}}\Big(1\!+\!\frac{7}{2} \mathcal{E}\!-\!\frac{4J}{m^2}\frac{ \mathcal{E}}{ \mathcal{J}}\Big)}\!=\!1\!-\!\Big[1\!+\!2 \mathcal{E} \mathcal{J}^2\!-\!\mathcal{E}(12\!+\!15 \mathcal{E} \mathcal{J}^2)\!+\!2\mathcal{E}\frac{q^2}{m^2}\!+\!\big(1\!+\! \mathcal{E}\mathcal{J}^2\big)\frac{16J\mathcal{E}}{m^2\mathcal{J}}\Big]^{\frac{1}{2}}\!\!\cos E.
%\end{eqnarray}
\begin{eqnarray}
&& \hskip -.5cm  r=-\frac{m }{2 \mathcal{E}}\Big(1\!+\!\frac{7}{2} \mathcal{E}\Big)\Big\{1\!-\!\Big[1\!+\!2 \mathcal{E} \mathcal{J}^2\!-\!\mathcal{E}(12\!+\!15 \mathcal{E} \mathcal{J}^2)\!+\!2\mathcal{E}\frac{q^2}{m^2}\Big]^{\frac{1}{2}}\!\cos E\Big\}~,
\end{eqnarray}
and the relation between the eccentric anomaly $E$ and the time $t$ can be written as
\begin{eqnarray}
&&  \hskip -1.5cm  t\,\Big(\frac{2\pi}{{\rm T}_{E}}\Big) = E - \Big[1+8\mathcal{E}\Big(1\!-\!\frac{7}{2}\mathcal{E}\Big)\Big]\Big[1\!+\!2 \mathcal{E} \mathcal{J}^2\!-\!\mathcal{E}(12\!+\!15 \mathcal{E} \mathcal{J}^2)\!+\!2\mathcal{E}\frac{q^2}{m^2}\Big]^{\frac{1}{2}} \sin E ~,\label{t-E2}
\end{eqnarray}
with the orbital period
\begin{eqnarray}
{\rm T}_E = {\rm T}_{\phi} = {\rm T}_{E'} = \frac{2\pi m}{(-2\mathcal{E})^{\frac{3}{2}}}\Big(1-\frac{15}{4}\mathcal{E}\Big)~.
\end{eqnarray}

\vskip 0.5cm

\section{Summary}\label{sec:5rd}
Based on the Reissner-Nordstr\"{o}m metric in the harmonic coordinates, we have derived the post-Keplerian solutions for the test particle's orbit and period in the Reissner-Nordstr\"{o}m spacetime via the Wagoner-Will representation, the Epstein-Haugan representation, the Brumberg representation and the Damour-Deruelle representation, respectively. Among these representations, the formalisms of Brumberg and Damour-Deruelle are in terms of the post-Keplerian parameters, while the formalisms of Wagoner-Will and Epstein-Haugan add the post-Newtonian effects from the source's mass and charge to the Kepler solutions directly. All these solutions are identical to the first PN order. For readers' reference, we also give the relations between the (post-)Keplerian parameters in these representations, as well as their relations to the orbital energy and angular momentum.

\vskip 0.5cm

\section*{Acknowledgements}
The author (W. Lin) would like to thank Prof. Remo Ruffini for inviting him to visit the ICRANet at Pescara where parts of this work were carried out. This research was supported in part by the National Natural Science Foundation of China (Grant Nos. 11647314 and 11847307).

%\appendix
\section*{References}

\bibliographystyle{spphys}       % APS-like style for physics
\bibliography{Reference_20190124}

\begin{thebibliography}{10}
\providecommand{\url}[1]{{#1}}
\providecommand{\urlprefix}{URL }
\expandafter\ifx\csname urlstyle\endcsname\relax
  \providecommand{\doi}[1]{DOI \discretionary{}{}{}#1}\else
  \providecommand{\doi}{DOI \discretionary{}{}{}\begingroup
  \urlstyle{rm}\Url}\fi

\bibitem{Hagihara1931}
Y.~Hagihara, Japan. J. Astron. Geophys. \textbf{8}, 67 (1931)

\bibitem{Gackstatter1983}
V.~Gackstatter, Ann. Phys. (Leipzig) \textbf{40}, 352 (1983)

\bibitem{GrunauKagramanova2011}
S.~Grunau, V.~Kagramanova, Phys. Rev. D \textbf{83}, 044009 (2011)

\bibitem{Mino2003}
Y.~Mino, Phys. Rev. D \textbf{67}, 084027 (2003)

\bibitem{FujitaHikida2009}
R.~Fujita, W.~Hikida, Class. Quantum Grav. \textbf{26}, 135002 (2009)

\bibitem{HackmannXu2013}
E.~Hackmann, H.~Xu, Phys. Rev. D \textbf{87}, 124030 (2013)

\bibitem{Shibata1994}
M.~Shibata, Phys. Rev. D \textbf{50}, 6297 (1994)

\bibitem{Ryan1996}
F.~Ryan, Phys. Rev. D \textbf{53}, 3064 (1996)

\bibitem{GPV1998a}
L.A. Gergely, Z.I. Perj\'{e}s, M.~Vas\'{u}th, Phys. Rev. D \textbf{57}, 876
  (1998)

\bibitem{Brumberg1972}
V.~Brumberg, \emph{Relativistic celesctial mechanics} (Nauka, Moscow in
  Russian, 1972)

\bibitem{WagonerWill1976}
R.V. Wagoner, C.M. Will, Astrophys. J. \textbf{210}, 764 (1976)

\bibitem{Epstein1977}
R.~Epstein, Astrophys. J. \textbf{219}, 92 (1977)

\bibitem{DamourDeruelle1985}
T.~Damour, N.~Deruelle, Ann. Inst. H. Poincar$\acute{e}$ \textbf{43}, 107
  (1985)

\bibitem{Haugan1985}
M.P. Haugan, Astrophys. J. \textbf{296}, 1 (1985)

\bibitem{Soffel1987}
M.H. Soffel, H.~Ruder, M.~Schneider, Celestial Mech. \textbf{40}, 77 (1987)

\bibitem{Soffel1989}
M.H. Soffel, \emph{Relativity in Astrometry, Celestial Mechanics and Geodesy}
  (Berlin: Springer, 1989)

\bibitem{KlionerKopeikin1994}
S.A. Klioner, S.M. Kopeikin, Astrophys. J. \textbf{427}, 951 (1994)

\bibitem{KopeikinEfroimskyKaplan2012}
S.M. Kopeikin, M.~Efroimsky, G.~Kaplan, \emph{Relativistic Celestical Mechanics
  of the Solar System} (Wiley-VCH, New York, 2012)

\bibitem{PoissonWill2014}
E.~Poisson, C.~Will, \emph{Gravity - Newtonian, Post-Newtonian, Relativistic}
  (Cambridge University Press, 2014)

\bibitem{DamourSchafer1988}
T.~Damour, G.~Sch\"{a}fer, Nuovo Cimento B \textbf{101}, 127 (1988)

\bibitem{SchaferWex1993}
G.~Sch\"{a}fer, N.~Wex, Phys. Lett. A \textbf{174}, 196 (1993)

\bibitem{MoraWill2004}
T.~Mora, C.M. Will, Phys. Rev. D \textbf{69}, 104021 (2004)

\bibitem{MGS2004}
R.M. Memmesheimer, A.~Gopakumar, G.~Sch\"{a}fer, Phys. Rev. D \textbf{70},
  104011 (2004)

\bibitem{Boetzel2017}
Y.~Boetzel, A.~Susobhanan, A.~Gopakumar, A.~Klein, P.~Jetzer, Phys. Rev. D
  \textbf{96}, 044011 (2017)

\bibitem{Wex1995}
N.~Wex, Class. Quantum Gravity \textbf{12}, 983 (1995)

\bibitem{KonigsdorfferGopakumar2005}
C.~K$\ddot{o}$nigsd$\ddot{o}$rffer, A.~Gopakumar, Phys. Rev. D \textbf{71},
  024039 (2005)

\bibitem{KMG2005}
Z.~Keresztes, B.~Mik\'{o}czi, L.A. Gergely, Phys. Rev. D \textbf{72}, 104022
  (2005)

\bibitem{THS2010}
M.~Tessmer, J.~Hartung, G.~Sh\"{a}fer, Class. Quantum Gravity \textbf{27},
  165005 (2010)

\bibitem{GopakumarSchafer2011}
A.~Gopakumar, G.~Sch\"{a}fer, Phys. Rev. D \textbf{84}, 124007 (2011)

\bibitem{BMFB2013}
A.~Boh\'{e}, S.~Marsat, G.~Faye, L.~Blanchet, Class. Quantum Gravity
  \textbf{30}, 075017 (2013)

\bibitem{GergelyKeresztes2015}
L.A. Gergely, Z.~Keresztes, Phys. Rev. D \textbf{91}, 024012 (2015)

\bibitem{Mikoczi2017}
B.~Mik\'{o}czi, Phys. Rev. D \textbf{95}, 064023 (2017)

\bibitem{Poisson1998}
E.~Poisson, Phys. Rev. D \textbf{57}, 5287 (1998)

\bibitem{BiniGeralico2014}
D.~Bini, A.~Geralico, Phys. Rev. D \textbf{89}, 044013 (2014)

\bibitem{LinJiang2014}
W.~Lin, C.~Jiang, Phys. Rev. D \textbf{89}, 087502 (2014)

\end{thebibliography}

\end{document}